\documentclass[%
aip,
amsmath,amssymb,
reprint,%
]{revtex4-1}

\usepackage{graphicx}
\usepackage{dcolumn}
\usepackage{bm}
\usepackage[utf8]{inputenc}
\usepackage[T1]{fontenc}
\usepackage{mathptmx}
\usepackage{etoolbox}
\usepackage{xcolor}
\usepackage{amssymb,amsfonts,amsmath}
\usepackage{float}
\usepackage{epstopdf}
\usepackage{xcolor}
\usepackage{textcmds}
\usepackage{textcomp}
\usepackage{soul}
\usepackage{csquotes}
\usepackage{amsbsy}
\usepackage{mathrsfs} 
\usepackage{bigints} 
\usepackage{amsthm} 
\usepackage[english]{babel}
\usepackage[shortlabels]{enumitem}
\usepackage{mathrsfs,bigints,mathtools,dsfont}
\usepackage[colorlinks=true,linkcolor=blue,citecolor=blue]{hyperref}%
\usepackage[toc]{appendix}

\makeatletter
\def\@email#1#2{%
	\endgroup
	\patchcmd{\titleblock@produce}
	{\frontmatter@RRAPformat}
	{\frontmatter@RRAPformat{\produce@RRAP{*#1\href{mailto:#2}{#2}}}\frontmatter@RRAPformat}
	{}{}
}%
\makeatother
\begin{document}

\title{Swarmalators on a ring with uncorrelated pinning}
	
	\author{Gourab Kumar Sar}
	\affiliation{Physics and Applied Mathematics Unit, Indian Statistical Institute, 203 B. T. Road, Kolkata 700108, India}
	
	\author{Kevin O'Keeffe}
	\affiliation{Senseable City Lab, Massachusetts Institute of Technology, Cambridge, MA 02139} 
	
	\author{Dibakar Ghosh}
	\affiliation{Physics and Applied Mathematics Unit, Indian Statistical Institute, 203 B. T. Road, Kolkata 700108, India}

	
	
	\thanks{Corresponding Auhtor: Dibakar Ghosh}
	\email{diba.ghosh@gmail.com (D Ghosh); mr.gksar@gmail.com (G.K.Sar)}

	\begin{abstract}
		\par {We present a case study of swarmalators (mobile oscillators) which move on a 1D ring and are subject to pinning. Previous work considered the special case where the pinning in space and the pinning in the phase dimension were correlated. Here we study the general case where the space and phase pinning are uncorrelated, both being chosen uniformly at random. This induces several new effects, such as pinned async, mixed states, and a first order phase transition. These phenomena may be found in real world swarmalators such as systems of vinegar eels, Janus matchsticks, electrorotated Quincke rollers or Japanese tree frogs.}
	\end{abstract}
	
	\maketitle

 	\begin{quotation}
		Pinning, the sticking of flow to defects in the underlying medium, is a central topic in statistical physics and nonlinear dynamics. 
It is key to the dynamics of charge density waves, for instance, which give rise to useful forms of nonlinear conduction \cite{balandin2021charge}. It is also the mechanism behind the flux pinning of type-II superconductors \cite{blatter1994vortices}. Here we study pinning in a new context: phase oscillators which swarm around in space and they synchronize in time.  The competition between the pinning, synchronization, and spatial self-assembly of such “swarmalators” gives rise to rich collective behavior which we investigate with a mix of numerics and analysis. 
	\end{quotation}

\section{Introduction}
Synchronization and swarming have been at the forefront of the research avenues while studying the collective behaviors of coupled systems. After the pioneering work of Kuramoto in 1975, studies on synchronization took a steep acceleration and have skyrocketed ever since~\cite{kuramoto1975international}. Such have been the impact of these studies that its application is ubiquitously found in technology from cardiac pacemaker cells~\cite{michaels1987mechanisms,verheijck1998pacemaker} to Josephson junction arrays~\cite{wiesenfeld1996synchronization} and power grid networks~\cite{motter2013spontaneous,dorfler2012synchronization} to name a few, apart from its overflowing examples in nature like flashing of fireflies~\cite{buck1976synchronous,buck1988synchronous}, chorusing frogs~\cite{aihara2008mathematical}, movement of crowds~\cite{ma2021spontaneous}, clapping in unison~\cite{neda2000physics} etc. On a parallel front, studies on swarming have also gained attention of the researchers and have not lagged far behind, especially after the seminal work of Vicsek et al.~\cite{vicsek1995novel}. Particularly, they have been given core attention when investigating the movements of animal groups, bird flocks, fish schools, insect swarms etc.~\cite{buhl2006disorder,couzin2007collective,ballerini2008interaction,sumpter2010collective,herbert2016understanding}. The understanding from these studies has been applied to technological fields like swarm robotics~\cite{ducatelle2011self,ducatelle2014cooperative}, aerial drones~\cite{tahir2019swarms,abdelkader2021aerial} and even for traffic controlling~\cite{ma2019green}.

Although there is quite a vast understanding of synchronization and swarming as independent fields, one can not say so when these two fields are brought into contact. The interplay between swarming and synchronization which is recently termed as \textit{swarmalation} lacks theoretical understanding. The co-occurrence of these two phenomena can be found in biological microswimmers~\cite{yang2008cooperation,riedel2005self,quillen2021metachronal,quillen2022fluid,peshkov2022synchronized,tamm1975role,verberck2022wavy,belovs2017synchronized}, driven colloids~\cite{yan2012linking,hwang2020cooperative,zhang2020reconfigurable,bricard2015emergent,zhang2021persistence,manna2021chemical,li2018spatiotemporal,chaudhary2014reconfigurable}, embryonic cells~\cite{tan2022odd,petrungaro2019information} etc. This necessitates careful and rigorous research on entities which swarm in space and sync over time, commonly known as \textit{swarmalators}.

Striding on that path, Tanaka et al. came up with a model of chemotactic oscillators where reaction among the particles were mediated by the surrounding chemicals\cite{tanaka2007general}. However, swarmalator as an entity representing the bidirectional interaction between spatial and internal dynamics was first introduced by O'Keeffe et al.~\cite{o2017oscillators} which paved the way for future works. This two dimensional (2D) model has been explored with various mechanisms thus far~\cite{o2018ring,ha2021mean,hong2021coupling,jimenez2020oscillatory,ha2019emergent,sar2022swarmalators,lizarraga2020synchronization,lee2021collective,degond2022topological,schilcher2021swarmalators,yadav2023exotic,ceron2023diverse,o2019review,sar2022dynamics,kongni2023phase,yadav2023exotic,ghosh2023antiphase}. Very recently a one dimensional (1D) counterpart of the 2D swarmalator model has been proposed by O'Keeffe et al.~\cite{o2022collective}. This model is essentially a pair of Kuramoto-like models which is analytically tractable and thus has captured the well deserved attention of the research community~\cite{yoon2022sync,o2022swarmalators,sar2023pinning,hong2023swarmalators,lizarraga2023synchronization,sar2023solvable,hao2023mixed,hong2023swarmalators}.

In this paper, we aim to study the dynamics of swarmalators on a 1D ring when random pinning is introduced in the system. Pinning generally refers to a substance's propensity to adhere to the irregularities of the underlying medium, necessitating external force to produce flow. Examples can be laid out in form of charge density waves~\cite{gor2012charge,gruner1985charge} or a pair of magnetic domain walls~\cite{hrabec2018velocity,haltz2021domain}. In the first example, the phases of the density waves which stand for $\theta_i$, located at positions $x_i$ in the underlying lattice, depin from their designated (pinned) values at a critical forcing amplitude. For the second case, the center of mass $x_i$ and the magnetic dipole vector $\theta_i$ of the magnetic domain walls overcome the pinned configuration and begin to interact when they are sufficiently forced. These implications motivate us to employ random pinning in systems consisting of position and phase interaction.

Swarmalators with spatial and phase pinning have been recently studied by Sar et al.~\cite{sar2023pinning} and diverse collective states including chaos and quasi-periodicity were reported. However, the spatial and phase pinning sites denoted by $\alpha_i$ and $\beta_i$, respectively were chosen linearly as $\alpha_i=\beta_i = 2 \pi i/N$, $N$ denoting the number of swarmalators. This form of pinning sites although helped in the analytical treatment, is unable to capture the broader scenario where the impurities of the underlying systems are more often than not randomly distributed~\cite{reichhardt2016depinning}. To fill this gap, here we have chosen both the spatial and phase pinning sites randomly and shown the robustness of the previously reported states as well as the emergence of new ones. We have also bailed ourselves out of the symmetric spatial and phase coupling strengths and explored the model when these strengths are not necessarily equal. Different phase transition phenomena have also been shown varying the driving/forcing amplitude and coupling strengths. The highlight is a first order phase transition from a pinned async state to a sync state. We believe our work adds to the existing results of swarmalators on disordered environment and hope that it injects impetus to further such explorations.

\section{Model Description}
The system we set out to study is a pair of Kuramoto-like models
\begin{align}
    \dot{x_i} &= E + b \sin(\alpha_i-x_i) + \frac{J}{N} \sum_j^N \sin(x_j - x_i) \cos(\theta_j - \theta_i), \label{eom-x} \\
    \dot{\theta_i} &= E +  b \sin(\beta_i-\theta_i) + \frac{K}{N} \sum_j^N \sin(\theta_j - \theta_i ) \cos(x_j - x_i ), \label{eom-theta}
\end{align}
\noindent
where $(x_i, \theta_i) \in (\mathbb{S}^1, \mathbb{S}^1)$ are the position and phase of the $i$-th swarmalator for $i=1,2,\ldots,N$ and $N$ is the number of swarmalators. $E$ represents external forcing or strength of driving. $J$, $K$ are the spatial and phase coupling strengths which control the phase dependent aggregation and position dependent synchronization, $b$ is the strength of the pinning. $\alpha_i$ and $\beta_i$ are the position and phase pinning sites, respectively which are chosen randomly from $[-\pi, \pi]$. $b$ $(>0)$ keeps the swarmalators pinned to their pinning sites, whereas, the coupling strengths $J$ $(>0)$ and $K$ $(>0)$ minimize the differences in both positions and phase. When $b \gg J,K$, swarmalators remain pinned and when $J,K \gg b$, they move close to each other and get synchronized in phase. The driving strength $E$ determines whether the solution will be static or moving (non-stationary). Without loss of generality, we set the pinning strength $b=1$ and study our model with three parameters $E$, $J$, and $K$. The $\alpha_i = \beta_i$ model with $J=K$ case has been previously studied in Ref.~\cite{sar2023pinning}. Here we focus on studying the model where the pinning locations are asymmetric, i.e., $\alpha_i \ne \beta_i$.  We explore both the cases when the coupling strengths are symmetric ($J=K$) and asymmetric ($J \ne K$).

We define the following order parameters which are instrumental for studying the collective states
\begin{align}
    W_{\pm} &= S_{\pm} e^{{i} \Phi_{\pm}} =  \frac{1}{N} \sum_j e^{{i}(x_j \pm \theta_j)}. \hspace{10pt} ({i} = \sqrt{-1})\label{order-par1}
\end{align}
$S_{\pm}$ measure the phase-space correlation of the swarmalators. If the phases and positions are perfectly correlated as $x_i = \pm \theta_i + C$ for some constant $C$, then we have $S_{\mp} = 1$. Otherwise $S_{\pm}<1$. In the complete asynchronous state where both the positions and phases are uniformly distributed between $-\pi$ to $\pi$, we get $S_{\pm} = 0$. We define two more order parameters to measure if the solutions are pinned or not,
\begin{align}
	Z_{x} &= R_{x} e^{{i} \Psi_{x}} =  \frac{1}{N} \sum_j e^{{i}(x_j - \alpha_j)}, \label{order-par2}\\
	Z_{\theta} &= R_{\theta} e^{{i} \Psi_{\theta}} =  \frac{1}{N} \sum_j e^{{i}(\theta_j - \beta_j)}. \label{order-par3}
\end{align}
By definition, $0<R_x,R_{\theta}<1$. When the positions of the swarmalators $x_i$ are pinned to the assigned locations $\alpha_i$, we get $R_x\approx 1$. Similarly, $R_{\theta} \approx 1$ when $\theta_i \rightarrow \beta_i$. We rewrite the model in terms of the order parameters $S_{\pm}$ as
\begin{align}
    \dot{x}_{i} &= E + \sin(\alpha_i-x_{i}) + \frac{J}{2}  S_+ \sin( \Phi_+ - (x_{i} + \theta_{i}) ) \nonumber \\ 
    \ &+ \frac{J}{2}  S_- \sin( \Phi_- - (x_{i} - \theta_{i}) ) , \\
    \dot{\theta}_{i} &= E + \sin(\beta_i-\theta_{i}) + \frac{K}{2} S_+ \sin( \Phi_+ - (x_{i} + \theta_{i}) ) \nonumber \\
    & - \frac{K}{2}  S_- \sin( \Phi_- - (x_{i} - \theta_{i}) ) .
\end{align}
Next, we investigate the emerging collective states of the model and their properties by changing three parameters $E$, $J$ and $K$.
\begin{figure*}[htpb!]
	\centering
	\includegraphics[width = 0.9\textwidth]{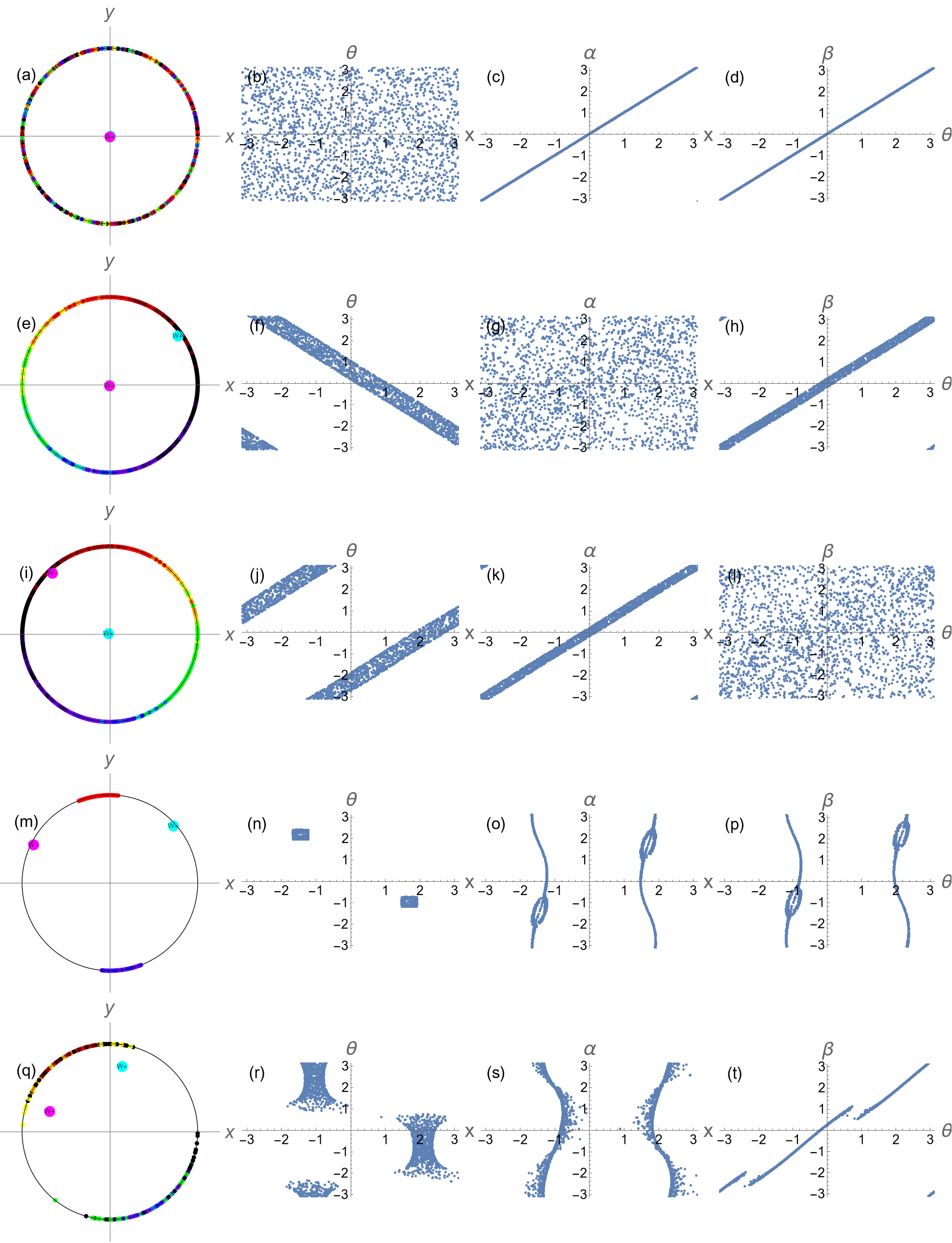}
	\caption{\textbf{Collective states in absence of driving ($E = 0$).} (a)-(d) Pinned async for $J =K= -1$. (e)-(h) Phase wave of type $x_i =-\theta_i +C$ for $J = -1, K = 5$. (i)-(l) Phase wave of type $x_i = \theta_i +C$ for $J = 5, K = -1$. (m)-(p) Sync state for $J = K=5$. (q)-(t) Mixed state for $J=5,K=0.5$. Here, $(dt,T,N) = (0.1, 200, 2000)$ and initially $x_i, \theta_i$ were drawn randomly from $[-\pi,\pi]$. The cyan and magenta circles in the first column refer to the complex order parameter $W_+$ and $W_-$, respectively. The pinning sites are also chosen randomly from $[-\pi,\pi]$. In the first column, the positions of the swarmalators on the ring are shown where they are colored according to their phases. The $x$-$\theta$, $x$-$\alpha$ and $\theta$-$\beta$ correlations are depicted in the second, third, and fourth columns, respectively for all the five states.}
	\label{states-undriven-1}
\end{figure*}
\section{Case I: $E=0$}
First, we set $E=0$, i.e., explore the collective states when there is no driving force present in the system. Numerics suggest four different collective states when the driving strength is zero and we vary the coupling strengths $J$ and $K$.\\

  1. \textit{Pinned async state}: Swarmalators' positions and phases are pinned to their pinning locations and we have $x_i \rightarrow \alpha_i$, $\theta_i \rightarrow \beta_i$ for all $i=1,2,\ldots N$. As a result $(R_x, R_{\theta}) = (1,1)$. Since the pinning locations are randomly distributed inside $[-\pi,\pi]$, the positions and phases of the swarmalators are uncorrelated in this state. Which gives $(S_+,S_-) = (0,0)$ here. See Fig.~\ref{states-undriven-1} first row for visual demonstration of this state. In Fig.~\ref{states-undriven-1}(a), the positions of the swarmalators on the 1D ring is delineated where they are colored according to their phases. Figure \ref{states-undriven-1}(b) shows the position-phase correlation in the $x$-$\theta$ plane. The relations between the positions and phases to the pinning sites are portrayed in Figs.~\ref{states-undriven-1}(c)-(d).\\
  
   2. \textit{Phase wave state}: In this state, the positions and phases of the swarmalators are found to be correlated. They can either be like $x_i= -\theta_i + C$ for some constant $C$ which yields $(S_+,S_-)=(1,0)$, or like $x_i=\theta_i + C$ where the order parameters $(S_+,S_-)=(0,1)$. These two instances of the phase wave state are depicted in the second and third columns of Fig.~\ref{states-undriven-1}, respectively. More generally, the phase wave state is characterized as the $(S,0)$ or $(0,S)$ state where $S \ne 0$. One of $R_x$ and $R_{\theta}$ takes value close to $1$ and the other is almost zero. The emergence of these two configurations of the phase wave state depends on the choice of initial conditions.\\
   
   3. \textit{Sync state}: As the name suggests, in here, we observe that the coupling strengths $J$ and $K$ overcome the pinned arrangement and swarmalators start to synchronize both their positions and phases. In Fig.~\ref{states-undriven-1}(m), they can be found forming two spatial clusters on the ring. The width of the clusters decreases with increment in the coupling strengths. The $x$-$\theta$ correlation in Fig.~\ref{states-undriven-1}(n) highlights the fact that both the positions and phases are synchronized here. Figures.~\ref{states-undriven-1}(o)-(p) depict the $x$-$\alpha$ and $\theta$-$\beta$ correlations, respectively. In terms of the order parameters, the sync state can be represented as $(S_+,S_-)=(S,S)$ where $S\ne 0$ and $R_x, R_{\theta} \ll 1$.\\
   
   4. \textit{Mixed state}: Alike in the sync state, here also both the order parameters $S_{\pm}$ are nonzero. But, they are not equal and that is why we call this as the mixed state. Order parameters characterize this state as $(S_+,S_-)=(S_1,S_2)$ where $S_1,S_2 \ne 0$ and $S_1 \ne S_2$. The fifth column of Fig.~\ref{states-undriven-1} illustrates this state. Either $x$ or $\theta$ are completely shifted from their pinning sites ($x$ in Fig.~\ref{states-undriven-1}(s)) and the other component is on the verge of losing the pinning configuration ($\theta$ in Fig.~\ref{states-undriven-1}(t)). As a result, one of $R_x$ and $R_{\theta}$ is much less than $1$ and the other is almost $1$.\\
    
   The system with no driving ($E=0$) is further studied under two scenarios. One, when the coupling strengths $J$ and $K$ are equal and two, when they are unequal.
    
\subsection{Symmetric coupling strengths ($J=K$)}
\begin{figure}[htp]
	\centering
	\includegraphics[width=\columnwidth]{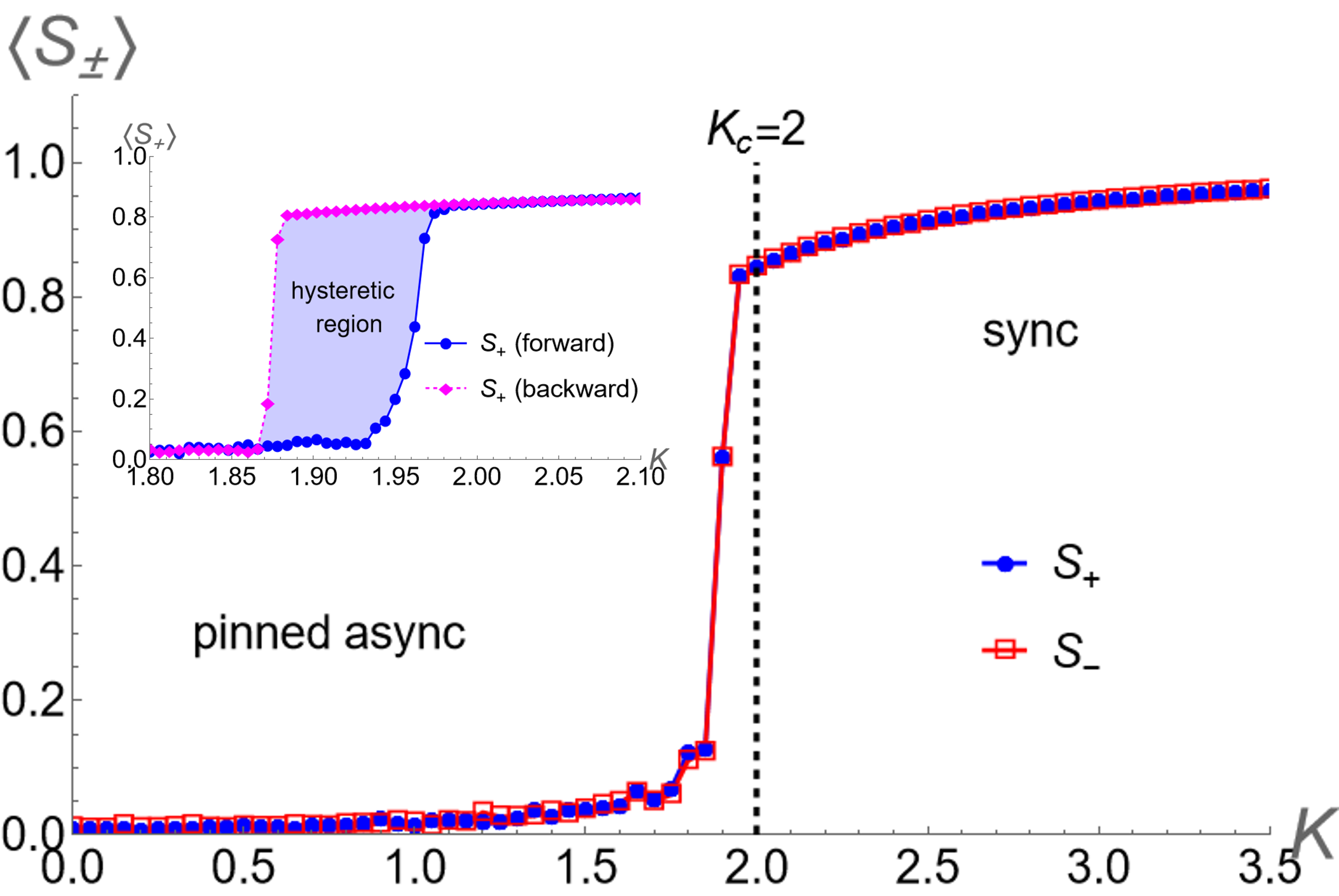}
	\caption{\textbf{Phase transition with symmetric coupling $J=K$ in absence of driving.} $S_+$ and $S_-$ are calculated as $K$ is being varied over an interval. $S_+$ (filled blue circles) and $S_-$ (void red squares) are equal in magnitude for all $K$. There is a transition from the $(S_+,S_-)= (0,0)$ (pinned async) state to the $(S,S)$ (sync) state near $K_c=2$. Simulation parameters are $(dt,T,N)=(0.5,200,10^4)$. In the inset, hysteretic phase transition is shown. Filled blue circles and magenta diamonds indicate the values of $S_+$ for forward and backward transitions, respectively. The shaded light blue region inside the forward and backward transition curves signify the hysteresis region. Simulation parameters: $(dt,T,N)=(0.5,200,10^5)$. Time average is taken over last $10\%$ data which is further averaged over $10$ realizations for all the data points.}
	\label{j=k-phase-transition}
\end{figure}

With symmetric coupling strengths, we find the occurrence of only those states where the values of the order parameters $S_+$ and $S_-$ are equal. These are the pinned async state and the sync state. Here, we are eventually left with only one parameter $K (=J)$ and we vary it to study the transition of states. In Fig.~\ref{j=k-phase-transition}, $S_+$ and $S_-$ are plotted as functions of $K$ by filled blue circles and void red square markers, respectively. They are joined by blue and red solid lines and we find that these lines overlap indicating the emergence of those states with $S_+=S_-$. From the figure we can see that there is a phase transition from the pinned async state where $(S_+,S_-)=(0,0)$ to the sync state $(S_+,S_-)=(S,S)$ ($S>0$) around $K=2$. The nature of the transition reveals that there is an abrupt jump of $S_{\pm}$ near this transition point. Which essentially instructs us to check for hysteresis which is associated with first order phase transition. In the inset of Fig.~\ref{j=k-phase-transition}, we plot the forward and backward transition curves for $S_+$ ($=S_-$) by filled blue circles and magenta diamonds, respectively over a zoomed in portion near the transition point $K_c=2$. In the forward transition, $S_+$ takes a sudden jump from the $(0,0)$ state to the $(S,S)$ state around $K \approx 1.97$. For the backward transition, the $(S,S)$ state persists till the value of $K$ drops down to $1.86$. The interval between these $K$ values observes the co-existence of the pinned async and the sync state resulting in a bistable hysteretic region.

Next, we proceed with the analytical treatment of the pinned async state. In the $N \rightarrow \infty$ limit, $x_{\alpha}$ is assigned to the pinning site $\alpha$ where we get rid of the suffix in $\alpha_i$. Similarly, $\theta_{\alpha}$ is assigned to the phase pinning site $\Gamma(\alpha)$ where $\Gamma$ is any map on $\mathbb{S}^1$. If $\Gamma(\alpha)=\alpha$ then we get the symmetric pinning. Under this set up, we reform the governing equations as
\begin{align}
    \dot{x}_{\alpha} &= E + \sin(\alpha-x_{\alpha}) + \frac{K}{2}  S_+ \sin( \Phi_+ - (x_{\alpha} + \theta_{\alpha}) ) \nonumber \\ 
    \ &+ \frac{K}{2}  S_- \sin( \Phi_- - (x_{\alpha} - \theta_{\alpha}) ) , \\
    \dot{\theta}_{\alpha} &= E + \sin(\Gamma(\alpha)-\theta_{\alpha}) + \frac{K}{2} S_+ \sin( \Phi_+ - (x_{\alpha} + \theta_{\alpha}) ) \nonumber \\
    & - \frac{K}{2}  S_- \sin( \Phi_- - (x_{\alpha} - \theta_{\alpha}) ) ,
\end{align}
where $S_{\pm}$ are rewritten as
\begin{equation}
    S_{\pm} e^{{i} \Phi_{\pm}} =  \frac{1}{2\pi} \int_{0}^{2\pi} e^{{i}(x_{\alpha} \pm \theta_{\alpha})} \, d\alpha .
    \label{order-par-inf}
\end{equation}
$x: \alpha \mapsto x_{\alpha}$ and $\theta: \alpha \mapsto \theta_{\alpha}$ are self-maps of the unit circle. In the async state, $S_{\pm}=0$ and the fixed points are of the form
\begin{align}
    x_{\alpha} &= \alpha + \sin^{-1}(E) \label{fpx},\\
    x_{\alpha} &= \Gamma(\alpha) + \sin^{-1}(E) \label{fpt}.
\end{align}
Here, we will study the local stability of the pinned async state. It can be achieved by diagonalizing the second variation of the potential function~\cite{strogatz1989collective} which is of the form
\begin{align}
    P(x_{\alpha},\theta_{\alpha}) = -E \int_0^{2\pi} x_{\alpha} \, d \alpha \; -E \int_0^{2\pi} \theta_{\alpha} \, d \alpha \nonumber \\ - \int_0^{2\pi} \cos(\alpha - x_{\alpha}) \, d \alpha - \int_0^{2\pi} \cos(\Gamma(\alpha) - \theta_{\alpha}) \, d \alpha  \nonumber \\ - \frac{K}{2\pi} \int_0^{2\pi} \int_0^{2\pi} \cos(x_{\beta} - x_{\alpha}) \cos(\theta_{\beta}-\theta_{\alpha}) \, d\alpha d \beta.
    \label{pot}
\end{align}
We consider a perturbation $\eta:\alpha \mapsto \eta_{\alpha}$ about the pinned async state
\begin{align}
    x_{\alpha}(\epsilon) &= \alpha + \sin^{-1}(E) + \epsilon \eta_{\alpha} ,\label{perturb1}\\
    \theta_{\alpha}(\epsilon) &= \Gamma(\alpha) + \sin^{-1}(E) + \epsilon \eta_{\alpha}.
    \label{perturb2}
\end{align}
Here $\epsilon$ is small in magnitude. We plug Eqs.~\eqref{perturb1}-\eqref{perturb2} into Eq.~\eqref{pot} and get
\begin{align}
    P(x_{\alpha}&(\epsilon),\theta_{\alpha}(\epsilon)) = -E \int_0^{2\pi} (\alpha + \sin^{-1}(E) + \epsilon \eta_{\alpha}) \, d \alpha \; \nonumber \\&-E \int_0^{2\pi} (\Gamma(\alpha) + \sin^{-1}(E) + \epsilon \eta_{\alpha}) \, d \alpha \; \nonumber \\ &- 2\int_0^{2\pi} \cos(\sin^{-1}(E) + \epsilon \eta_{\alpha}) \, d \alpha \nonumber \\ &- \frac{K}{4\pi} \int_0^{2\pi} \int_0^{2\pi} \cos(\beta - \alpha +\epsilon \eta_{\beta} - \epsilon \eta_{\alpha}) \nonumber \\ &\qquad \times \cos(\Gamma(\beta) - \Gamma(\alpha) +\epsilon \eta_{\beta} - \epsilon \eta_{\alpha}) \, d\alpha d \beta.
    \label{pot2}
\end{align}
We calculate the second variation of $P$ which is a quadratic form
\begin{equation}
    \Gamma(\eta) = \frac{d^2}{d \epsilon^2} P(x_{\alpha}(\epsilon))\Big |_{\epsilon=0}.
\end{equation}
After performing the differentiation, we get
\begin{align}
    \Omega(&\eta) = 2 \sqrt{1-E^2} \int_{0}^{2\pi} \eta^2_{\alpha} \, d\alpha \nonumber \\
    &+ \frac{K}{2\pi} \int_{0}^{2\pi} \int_{0}^{2\pi} (\eta_{\beta} - \eta_{\alpha})^2 \cos(\gamma(\beta) - \gamma(\alpha)) \, d \alpha d \beta ,
    \label{sec-var}
\end{align}
where $\gamma(\alpha) = \alpha + \Gamma(\alpha)$. We work with $\Gamma(\alpha)=(r-1) \alpha$, $r \in \mathbb{Z}$ so that $\gamma(\alpha) = r \alpha$.
The $(\eta_{\beta} - \eta_{\alpha})^2$ term on the RHS of Eq.~\eqref{sec-var} is expanded the and further simplified as
\begin{align}
    \int_{0}^{2\pi} &\int_{0}^{2\pi} (\eta_{\beta} - \eta_{\alpha})^2 \cos(r\beta - r \alpha) \, d \alpha d \beta \nonumber \\ &= -2 \int_{0}^{2\pi} \int_{0}^{2\pi} \eta_{\beta} \eta_{\alpha} \cos(r \beta - r \alpha) \, d \alpha d \beta .
    \label{eq43}
\end{align}
Now, the $\alpha$ and $\beta$ integrals separate on expanding the term $\cos(r\beta - r \alpha)$:
\begin{align}
    &\int_{0}^{2\pi} \int_{0}^{2\pi} \eta_{\beta} \eta_{\alpha} \cos(r\beta - r \alpha) \, d \alpha d \beta \nonumber \\ &= \int_{0}^{2\pi} \int_{0}^{2\pi} \eta_{\beta} \eta_{\alpha} (\cos{r\alpha} \cos{r\beta} + \sin r \alpha \sin r \beta) \, d \alpha d \beta \nonumber \\ &= \Bigg[ \int_{0}^{2\pi} \eta_{\alpha} \cos r \alpha \, d\alpha\Bigg]^2 + \Bigg[ \int_{0}^{2\pi} \eta_{\alpha} \sin r \alpha \, d\alpha\Bigg]^2 \nonumber \\ &= \Bigg| \int_{0}^{2\pi} \eta_{\alpha} e^{{i} r \alpha} \, d\alpha \Bigg|^2 = 4 \pi^2 \big|\hat{\eta}_{\alpha}(-r)\big|^2 ,
    \label{eq44}
\end{align}
where $\hat{\eta}_{\alpha}$ denotes the Fourier transform of $\eta_{\alpha}$ which is defined by
\begin{equation}
    \hat{\eta}_{\alpha}(m) = \frac{1}{2\pi} \int_{0}^{2\pi} \eta_{\alpha} e^{-{i} m \alpha} \, d\alpha.
\end{equation}
We move to the Hilbert space $L^2(\mathbb{S}^1)$ with the inner product defined on it as,
\begin{equation}
    \mu_{\alpha} \cdot \nu_{\alpha} = \frac{1}{2\pi} \int_{0}^{2\pi} \mu_{\alpha} \nu_{\alpha} \, d\alpha .
\end{equation}
Let us choose $\mu_{\alpha} = \cos r \alpha$, $\nu_{\alpha} = \sin r \alpha$ and we have $||\mu_{\alpha}||^2 = ||\nu_{\alpha}||^2 = 1/2$ and $\mu_{\alpha} \cdot \nu_{\alpha} = 0$. We consider $\eta_{\alpha}^{\perp}$, a function which is orthogonal to both $\mu_{\alpha}$ and $\nu_{\alpha}$. This means $\mu_{\alpha} \cdot \eta_{\alpha}^{\perp} = \nu_{\alpha} \cdot \eta_{\alpha}^{\perp} = 0$. Now, $\eta_{\alpha}$ can be expressed as a linear combination of $\mu_{\alpha}$, $\nu_{\alpha}$, $\eta_{\alpha}^{\perp}$ as:
\begin{equation}
    \eta_{\alpha} = p \frac{\mu_{\alpha}}{||\mu_{\alpha}||} + q \frac{\nu_{\alpha}}{||\nu_{\alpha}||} + \eta_{\alpha}^{\perp},
\end{equation}
Also, we get
\begin{equation}
    ||\eta_{\alpha}||^2 = p^2 + q^2 + ||\eta_{\alpha}^{\perp}||^2.
    \label{eq48}
\end{equation} 
With these tools, it is elementary to see that
\begin{align}
    \big|\hat{\eta}_{\alpha}(-r)\big|^2 &= \Bigg| \frac{1}{2\pi} \int_{0}^{2\pi} \eta_{\alpha} e^{{i} r \alpha} \, d\alpha \Bigg|^2 \nonumber \\ &= \Bigg[ \int_{0}^{2\pi} \eta_{\alpha} \cos r \alpha \, d\alpha\Bigg]^2 + \Bigg[ \int_{0}^{2\pi} \eta_{\alpha} \sin r \alpha \, d\alpha \Bigg]^2 \nonumber \\ &= (\eta_{\alpha} \cdot \mu_{\alpha})^2 + (\eta_{\alpha} \cdot \nu_{\alpha})^2 \nonumber \\
    &= (p||\mu_{\alpha}||)^2 + (q||\nu_{\alpha}||)^2 = \frac{p^2+q^2}{2}.
    \label{eq49}
\end{align}
After substitutions and simplification, from Eq.~\eqref{sec-var} we finally get
\begin{align}
    &\Omega(\eta) \nonumber \\ &= 4\pi \Bigg[ \sqrt{1-E^2}(p^2+q^2+||\eta_{\alpha}^{\perp}||^2) - K\Big(\frac{p^2+q^2}{2}\Big) \Bigg] \nonumber \\ &= 4\pi \Bigg[ (p^2+q^2) \Big( \sqrt{1-E^2} - \frac{K}{2}\Big) + \sqrt{1-E^2} ||\eta_{\alpha}^{\perp}||^2 \Bigg].
    \label{eq50}
\end{align}
So, $\Omega$ is positive definite when $\sqrt{1-E^2} - K/2 >0$. From this we get the stability boundary
\begin{align}
    E_c(K) =  \sqrt{1 - \frac{K^2}{4}} . \label{critical}
\end{align}
When $E=0$, we get $K_c=2$ which matches the transition point found in Fig.~\ref{j=k-phase-transition}. The small deviation from the analytically predicted value is due to the finite system size and reduces with increasing $N$.
\subsection{Asymmetric coupling strengths ($J \ne K$)}
\begin{figure}
	\centering
	\includegraphics[width=\columnwidth]{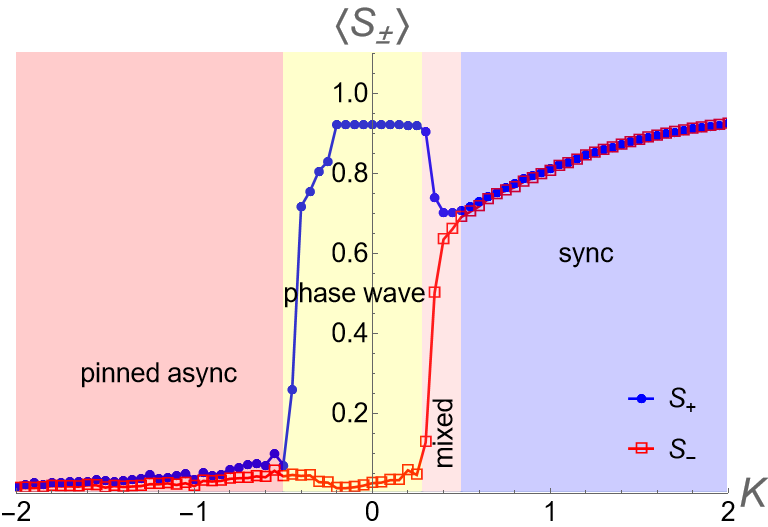}
	\caption{\textbf{Phase transition scenario with asymmetric coupling strengths $J \ne K$ for $E=0$.} Filled blue circles and void red markers represent the values of $S_+$ and $S_-$, respectively. The regions corresponding to the emergence of different states are colored by light red (pinned async), light yellow (phase wave), light pink (mixed) and light blue (sync). Simulation parameters used here are $(dt,T,N)=(0.1,200,10^4)$. Both $S_+$ and $S_-$ are time averaged over last $10\%$ data and further averaged over $10$ realizations. We have fixed $J=4$ here. The phase transition here follows the route pinned async $(0,0)$ $\rightarrow$ phase wave $(S,0)$ $\rightarrow$ mixed $(S_1,S_2)$ $\rightarrow$ sync $(S,S)$ with increasing $K$.}
	\label{phase-transition-j-ne-k-e-0}
\end{figure}

We found only two collective states out of the four mentioned at the beginning of the section with symmetric coupling strengths $J=K$. The rest of the two states which are the phase wave and mixed state can only be seen when $J$ and $K$ are unequal. For a particular case, we fix $J=4$ and inspect the different phase transition behaviors while varying $K$. The result is displayed in Fig.~\ref{phase-transition-j-ne-k-e-0}. Starting from the left with negative $K$, first we encounter the pinned async state where both $S_+$ and $S_-$ lie near zero. This is highlighted by the light red region in Fig.~\ref{phase-transition-j-ne-k-e-0}. Moving to the right with increasing $K$, we find that pinned async state becomes unstable near $K \approx -0.5$ where $S_+$ starts to increase from zero while $S_-$ continue to stay near zero (look at the light yellow region in Fig.~\ref{phase-transition-j-ne-k-e-0}). This is the $(S,0)$ state or the phase wave as we have discussed earlier. It is to be noted that, the phase wave state in this region can be either of the $(S,0)$ or $(0,S)$ types depending on the initial condition. We set the maximum of $S_+$ and $S_-$ to $S_+$ so that it is always of type $(S,0)$ during numerical simulation. With increasing $K$, the phase wave state loses its stability around $K \approx 0.3$. From the phase wave state, $S_+$ decreases and $S_-$ increases resulting in the occurrence of the $(S_1,S_2)$ case where $S_1>S_2$. This is the mixed state which is indicated by the light pink region in Fig.~\ref{phase-transition-j-ne-k-e-0}. From here, $S_+$ further decreases while $S_-$ increases until they collide around $K \approx 0.5$. Finally, we observe the sync state characterized by $(S,S)$ where $S \ne 0$. The interval of occurrence of the sync state is delineated by the light blue region in Fig.~\ref{phase-transition-j-ne-k-e-0}.

\begin{figure}[htpb!]
	\centering
	\includegraphics[width = \columnwidth]{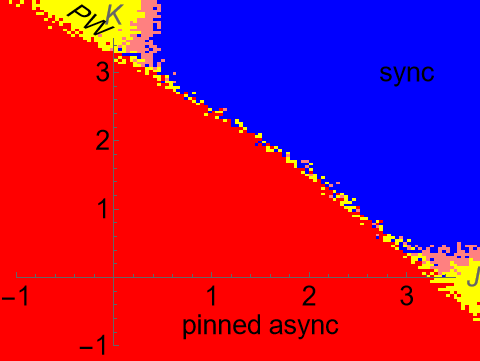}
	\caption{\textbf{Phase diagram in the $J$-$K$ plane for $E=0$}.  Red, yellow, pink and blue indicate the pinned async, phase wave (PW), mixed and sync states, respectively. We use the following simulation parameters: $(dt,T,N) = (0.1, 500, 1000)$. First $90\%$ data were discarded as transients. Each data point is an average of $5$ realizations. We have characterized the states in terms of the order parameters $S_{\pm}$. It can be seen that one of $J$ and $K$ has to be positively large for the phase wave state to be achieved.}
	\label{E-0-J-K-space}
\end{figure}

We have also looked for the bifurcation in the $J$-$K$ parameter plane. We demonstrate the diagram in Fig.~\ref{E-0-J-K-space}. We numerically calculate this by dividing the plane into $100$ by $100$ mesh points. At each of these point, the values of the order parameters $S_+$, $S_-$ are calculated. We color the mesh points by red, yellow, pink and blue if the emerging states at those points are pinned async, phase wave, mixed and sync, respectively. We have distinguished the states as below
\begin{align}
    & \text{Pinned async}: \{S_+<0.2\} \cap \{S_-<0.2\}. \nonumber \\
    & \text{Phase wave}: \{S_+>0.2\} \cap \{S_-<0.2\}. \nonumber\\
    & \text{Mixed}: \{S_+>0.2\} \cap \{S_->0.2\} \cap  \{S_+-S_->0.1\}. \nonumber \\
    & \text{Sync}: \{S_+>0.2\} \cap \{S_->0.2\} \cap  \{S_+-S_-<0.1\}. \nonumber
\end{align}
Based on the above conditions, we compute the bifurcation diagram in Fig.~\ref{E-0-J-K-space}. We find the existence of pinned async state (red region) when $J+K \lessapprox 3$. The phase wave state (yellow region) is encountered when one of $J$ and $K$ is positively large and the other is small. The mixed state (pink region) is seen at the boundary between the phase wave and the sync state and their region of occurrence is small compared to the other states. When both the coupling strengths $J$ and $K$ are positive and large, the sync state (blue region) prevails.

\section{Case II: $E \ne 0$}
With a sound understanding of the collective states in the $E=0$ limit, we now move to explore the dynamical features of the system in presence of driving, i.e., in the $E \ne 0$ limit. In the undriven limit, all the states were stationary. The mean or average velocity $\langle V \rangle = N^{-1} \sum_i \sqrt{ v_{x,i}^2 + v_{\theta,i}^2 } $ was zero there. But, in presence of driving in the system, we expect moving solutions where $\langle V \rangle \ne 0$. Looking at the fixed points given by Eqs.~\eqref{fpx}-~\eqref{fpt}, we can say that they exist when $E<1$. From numerics, we observe that all the solutions for $E>1$ are nonstationary in nature, i.e., $\langle V \rangle >0$.

\par
\begin{figure}[htpb!]
	\centering
	\includegraphics[width = \columnwidth]{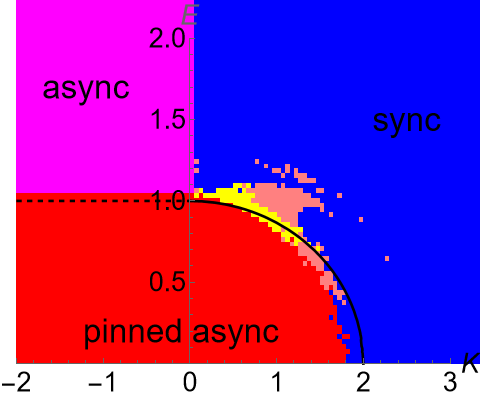}
	\caption{\textbf{Phase diagram in the $K$-$E$ plane for $J=K$}. Red, magenta, yellow, pink and blue colored regions correspond to the occurrence of pinned async, async, phase wave, mixed and sync states, respectively. Black dashed line is the $E=1$ line. Solid black line is the analytical prediction Eq.~\eqref{critical}. Here, we have used $(dt,T,N) = (0.1, 500, 1000)$. First $90\%$ data were discarded as transients. Each data point is an average of $5$ realizations. We see that with negative $K$ the pinned async state becomes unstable as $E$ crosses 1 from below and the nonstationary async state takes over. The phase wave and mixed states are observed over relatively small regions near the boundary between the pinned async and sync states.}
	\label{E-ne-0-K-E-space}
\end{figure}

 Another stark contrast with the $E=0$ case is that even with the symmetric coupling strengths $J=K$ we find the existence of phase wave and mixed states. Moreover, a new state is born where the positions and phases of the swarmalators are distributed inside $[-\pi,\pi]$ but unlike in the pinned async state they lose their connection with the pinning locations. In terms of the order parameter, in this state we have $S_{\pm} \approx 0$ and $R_x,R_{\theta} \ll 1$. This state is called as the \textit{async} state. To capture the entire scenario, we move to the $K$-$E$ parameter space. Figure~\ref{E-ne-0-K-E-space} captures the overall behavior of the system in the undriven limit with symmetric coupling strengths. Along with the aforementioned conditions to distinguish the pinned async, phase wave, mixed and sync states, here we write down the condition for realization of the async state as
\begin{align}
    & \text{Async}: \{S_+<0.2\} \cap \{S_-<0.2\} \cap \{R_x<0.2\} \cap \{R_{\theta}<0.2\}. \nonumber 
\end{align}
In Fig.~\ref{E-ne-0-K-E-space}, the black dashed line indicates $E=1$ beyond which the stationary pinned async state (red region) loses its stability and the nonstationary async state (magenta region) takes place. The solid black line signifies the analytical boundary Eq.~\eqref{critical} between the pinned async state and the sync state (blue region). We also observe the scattered presence of the phase wave (yellow) and the mixed (pink) states.

\begin{figure*}
    \centering
    \includegraphics[width=2\columnwidth]{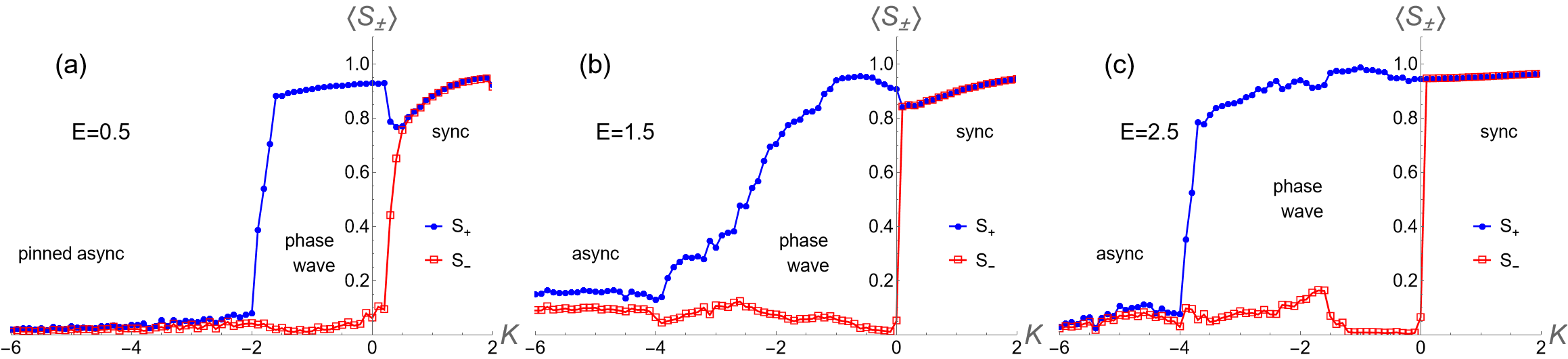}
    \caption{\textbf{Phase transition with asymmetric coupling $J \ne K$ with $E \ne 0$.} Filled blue circles and void red squares correspond to the values of $S_+$ and $S_-$, respectively. The transition phenomena is shown in (a) for $E=0.5$, in (b) for $E=1.5$ and in (c) for $E=2.5$. Simulation parameters: $(dt,T,N)=(0.5,200,10^4)$. Both the order parameters are time averaged over last $10\%$ data and averaged over $5$ realizations. Here, we have used $J=4.0$. In (a), we find pinned async $\rightarrow$ phase wave $\rightarrow$ mixed $\rightarrow$ sync, whereas in (b) and (c) we observe async $\rightarrow$ phase wave $\rightarrow$ sync. We also observe first-order transition taking place from phase wave state to sync state when $E$ is large.}
    \label{E-ne-0-j-ne-k}
\end{figure*}

\par We continue our study with asymmetric coupling strengths $J \ne K$. Again for instance we fix $J=4$. We accumulate the transition phenomena for different $E$ values in Fig.~\ref{E-ne-0-j-ne-k} where we vary $K$ over an interval $[-6,2]$. For $E=0.5$ in Fig.~\ref{E-ne-0-j-ne-k}(a), the transition behavior is similar to the one we found in Fig.~\ref{phase-transition-j-ne-k-e-0} with $E=0$. The route with increasing $K$ is noted down as: pinned async $\rightarrow$ phase wave $\rightarrow$ mixed $\rightarrow$ sync. An increase to the value of $E$ to $1.5$ shows the disappearance of the stationary pinned async state and arrival of the nonstationary async state in place of it. Moreover, in Fig.~\ref{E-ne-0-j-ne-k}(b) we observe that the interval for phase wave state expands compared to the $E=0.5$ case in Fig.~\ref{E-ne-0-j-ne-k}(a). The mixed state which exists over a tiny interval for $E=0.5$ vanishes completely for $E=1.5$. As a result, the transition route is: async $\rightarrow$ phase wave $\rightarrow$ sync. With $E=2.5$ in Fig.~\ref{E-ne-0-j-ne-k}(c), the dynamical features do not change much from the $E=1.5$ scenario. However, the phase wave state is more prominent here. We also see that with higher driving strength as the mixed state vanishes, the transition from the phase wave state to the sync state becomes abrupt.

\section{Discussion}
In this work, we have broken the extreme symmetries of the pinning model by taking both the pinning locations randomly. This resulted in the emergence of a pinned async state where $x_i \rightarrow \alpha_i$ and $\theta_i \rightarrow \beta_i$ which trivially transforms into the phase wave state if we assume $\alpha_i = \beta_i$. On top of it, with the asymmetric random pinning sites, we encounter phase wave states that arise when one of the position and phase variables remains pinned and the other depins. Under sufficiently large coupling strengths, sync state is observed where positions and phases of the swarmalators are synchronized. At an intermediate stage between the phase wave and sync state, the mixed state develops for small $E$ and disappears when $E$ is large.

Randomness is always difficult to deal with while treating it analytically compared to systematic patterns. While the random pinning locations make our model realistic and produce unknown collective states, they take away from us the luxury of deterministic mathematical framework. That is why we had to keep ourselves limited to numerical treatment of majority of the collective states. This being said, the collective states along with their counterparts in real world have enough to offer to this line of study.

Our model captures the behavior of real world swarmalators like circularly confined spermatazoa \cite{creppy2016symmetry} and bordertaxic vinegar eels \cite{quillen2021metachronal,quillen2022fluid,peshkov2022synchronized}. When combined with the findings of symmetric pinning~\cite{sar2023pinning}, the model's dynamical states can analogously be found in models of Japanese tree frogs~\cite{aihara2014spatio} and Janus matchsticks ~\cite{chaudhary2014reconfigurable}. The 1D ring model also has implications in groups of sperm~\cite{creppy2016symmetry} and forced colloids~\cite{yan2012linking,yan2015rotating,zhang2020reconfigurable}.

In a nutshell, our model establishes dynamics of swarmalators with random pinning locations with both symmetric and asymmetric coupling strengths. Future research works can look for an analytical treatment of these states in terms of the variation of order parameters, $\dot{S}_{\pm}$. The driving strength $E$ can also be chosen asymmetrically in the position and phase components, so is the pinning strength $b$. Local coupling can be introduced in the model. This work can also be extended to swarmalators where the spatial movements are in the two dimensional plane. The coupling strengths, driving force and the pinning strength which are identical (i.e., chosen from a delta distribution) here for all the swarmalators, can be chosen from some other kind of distributions as well. Periodic forcing/driving strengths can also be considered in place of the constant driving.

\color{black}
    
\section*{\label{sec:level6}Data Availability}
The data that support the findings of this study are available
from the corresponding author upon reasonable request.
%

\end{document}